%% file: main.tex
\documentclass[aps,prc,twocolumn,nofootinbib,superscriptaddress]{revtex4-1}

\usepackage{bm}        
\usepackage{xcolor}

\usepackage{dcolumn}   
\usepackage{bm}        
\usepackage{multirow}

\usepackage{amsmath}
\usepackage{amsfonts}
\usepackage{amssymb}
\usepackage{makeidx}
\usepackage{graphicx}
\usepackage{anysize}
\usepackage{hyperref}
\usepackage{slashed}
 \usepackage{bbm}

\usepackage{cleveref}

\usepackage{xcolor}

\usepackage{ulem} 

\newenvironment{mymathbox}
{\par\smallskip\centering\begin{lrbox}{0}%
\begin{minipage}[c]{0.8\textwidth}}
{\end{minipage}\end{lrbox}%
\framebox[0.9\textwidth]{\usebox{0}}%
\par\medskip
\ignorespacesafterend}
\newcommand{\bb}{\begin{mymathbox}}
\newcommand{\eb}{\end{mymathbox}}

\newcommand{\bs}{\boldsymbol}

\newcommand{\be}{\begin{equation}}
\newcommand{\ee}{\end{equation}}
\newcommand{\ba}{\begin{eqnarray}}
\newcommand{\ea}{\end{eqnarray}}

\newcommand{\np}{{\bf      p}}

\newcommand{\npsi}{{\bf \npsi}}

\newcommand{\Psib}{\overline{\Psi}}

\newcommand{\non}{\nonumber}

\newcommand{\bma}{\begin{pmatrix}}
\newcommand{\ema}{\end{pmatrix}}

\begin{document}

\title{Relativistic two-body currents for one-nucleon knockout in electron-nucleus scattering}

\author{T. Franco-Munoz}
\affiliation{Grupo de F\'isica Nuclear, Departamento de Estructura de la Materia, F\'isica T\'ermica y Electr\'onica, Facultad de Ciencias F\'isicas, Universidad Complutense de Madrid and IPARCOS, CEI Moncloa, Madrid 28040, Spain}
\author{J.~Garc\'ia-Marcos}
\affiliation{Grupo de F\'isica Nuclear, Departamento de Estructura de la Materia, F\'isica T\'ermica y Electr\'onica, Facultad de Ciencias F\'isicas, Universidad Complutense de Madrid and IPARCOS, CEI Moncloa, Madrid 28040, Spain}
\affiliation{Department of Physics and Astronomy, Ghent University, B-9000 Gent, Belgium}
\author{R.~Gonz\'alez-Jim\'enez}
\affiliation{Grupo de F\'isica Nuclear, Departamento de Estructura de la Materia, F\'isica T\'ermica y Electr\'onica, Facultad de Ciencias F\'isicas, Universidad Complutense de Madrid and IPARCOS, CEI Moncloa, Madrid 28040, Spain}
\author{J.M.~Ud\'ias}
\affiliation{Grupo de F\'isica Nuclear, Departamento de Estructura de la Materia, F\'isica T\'ermica y Electr\'onica, Facultad de Ciencias F\'isicas, Universidad Complutense de Madrid and IPARCOS, CEI Moncloa, Madrid 28040, Spain}

\date{\today}

\begin{abstract}
We present a detailed study of the contribution from two-body currents to the one-nucleon knockout process induced by electromagnetic interaction. The framework is a relativistic mean-field model (RMF) in which bound and scattering nucleons are consistently described as solutions of Dirac equation with potentials. We show results obtained with the most general expression of the two-body operator, in which the intermediate nucleons are described by relativistic mean-field bound states; then, we propose two approximations consisting in describing the intermediate states as nucleons in a relativistic Fermi gas, preserving the complexity and consistency in the initial and final states. These approximations simplify the calculations considerably, allowing us to provide outcomes in a reasonable computational time. The results obtained under these approximations are validated by comparing with those from the full model. Additionally, the theoretical predictions are compared with experimental data of the longitudinal and transverse responses of carbon 12. The agreement with data is outstanding for the longitudinal response, where the contribution from the two-body operator is negligible. In the transverse sector, the two-body current increases the response from 30 to 15\%, depending on the approximations and kinematics, in general, improving the agreement with data.    
\end{abstract}

\maketitle

\input{1_introduction}

\input{2_themodel}

\input{3_results}

\input{4_conclusions}

\begin{acknowledgments}
This work was supported by the Madrid Government under the Multiannual Agreement with Complutense University in the line Program to Stimulate Research for Young Doctors in the context of the V PRICIT (Regional Programme of Research and Technological Innovation), project PR65/19-22430; by 
project PID2021-127098NA-I00 funded by MCIN/AEI/10.13039/501100011033/FEDER, UE;
and by project RTI2018-098868-B-100 (MCIN/AEI,FEDER,EU). 
The computations of this work were performed in Brigit, the HPC server of the Complutense University of Madrid.
\end{acknowledgments}

\bibliography{bibliography}

\end{document}

%% file: 1_introduction.tex
\section{Introduction}

New generation of accelerator-based neutrino oscillation experiments, DUNE \cite{DUNE16} and T2HK \cite{T2K20}, require an unprecedent level of accuracy to succeed in determining neutrino properties as the CP-violating phase in the lepton sector and the neutrino-mass hierarchy. However, a major source of systematic uncertainty comes from the modeling of neutrino-nucleus interaction.  Therefore, an accurate description of these reactions has become one of the top challenges for theoretical nuclear physics \cite{Alvarez-Ruso18}.

Electron scattering is a powerful tool to obtain information about the response of nuclei and, due to its connection with neutrino-nucleus interaction, the interest on it has been renewed \cite{Amaro20,Katori17,Gu21,Khachatryan21}. In contrast with neutrino beams, where the incident neutrino energy is not a priori known and only flux distributions are available, electron scattering has the advantage of nearly monochromatic beams with well determined energies. In this way, the comparison with electron experimental data provides a benchmark for the validation of theoretical nuclear models. Hence, before applying any model in neutrino oscillation analyses, it is mandatory to first validate it against electron data. However, such a test is necessary but not sufficient to ensure the validity of a model, since only the vector part of the weak response can be tested through the electromagnetic response.

In the energy regime of accelerator-based neutrino oscillation experiments, the neutrino-nucleus interaction is driven by several reaction mechanisms: elastic scattering, discrete and collective nuclear excitations, quasielastic (QE) scattering, multinucleon knockout processes, pion production, and deep-inelastic scattering. Our focus is placed on the quasielastic region, where the incoming neutrino scatters off a nucleon, bound by the nuclear potential. QE scattering is the main interaction mechanism for neutrinos with energies around 1 GeV, being key to the understanding of neutrino interactions with nuclei and, consequently, its properties. In this work, we explore the impact of pions on the particle-hole (1p-1h) excitations, considering the interaction between nucleons through one pion exchange as two-body meson-exchange currents (MEC). It includes the contribution from the Delta-resonance mechanisms (Fig.~\ref{fig:delta}) and the contributions deduced from the chiral perturbation theory Lagrangian of the pion-nucleon system \cite{Scherer12} (Fig.~\ref{fig:background}, ChPT background or, simply, background terms in what follows). We construct a two-body current containing all Feynman diagrams with one exchanged pion between two nucleons through the mentioned mechanisms that results in a 1p-1h excitation. 

Here, we present a fully relativistic and quantum mechanical model for the simultaneous computation of the 1p-1h longitudinal and transverse electromagnetic nuclear responses. We consider an independent particle shell model approach where the initial particle state is described by a relativistic mean-field (RMF) model, meanwhile the final state is computed as a solution of the Dirac equation in the continuous with the energy-dependent relativistic mean-field (ED-RMF) potential. The key contribution of this work is the introduction of two-body meson-exchange currents. Particle-hole excitations through two-body currents present an intermediate bound-nucleon state which, in the most general approach, is described as a bound particle by the RMF potential. However, the computational effort that require these calculations motivated us to introduce two approximations: the description of the intermediate state as free Dirac spinors in a relativistic Fermi gas (RFG) and its extension including modified mass and energy that account for the relativistic interaction of nucleons. Our results for the $^{12}$C inclusive electromagnetic response functions including two-body currents show an increase of the transverse response, improving the agreement with data, meanwhile in the longitudinal one the effect is hardly visible.

Our results agree with recent calculations based on the non-relativistic \textit{ab initio} model of \cite{Lovato16}, with an increase of the transverse response over the whole energy transfer spectrum and little effect in the longitudinal sector. This is also obtained with the relativistic models of Refs.~\cite{Dekker94,Umino95b,Amaro02}, in which the two-body currents were studied within a relativistic Fermi gas approach.

The paper is organized as follows. In Section \ref{sec:model} we present our general formalism for the treatment of the nuclear structure. Then, in Section \ref{sec:two-body} we introduce the two-body contributions in the general case and, later, with the two possible approximations. Results for the inclusive electromagnetic responses of $^{12}$C are shown and analyzed in Section \ref{sec:results}.  Finally, in Section \ref{sec:conclusions} we  draw our conclusions.

%% file: 2_themodel.tex
\section{Nuclear model}\label{sec:model}

We consider that only one boson is exchanged between the leptons and hadron system. The impulse approximation (IA), in which one considers that the boson couples only the knocked out nucleon, gives the main contribution to the QE response, through the one-body current. In this work, we go beyond IA and include two-body currents within a fully relativistic framework.  

In our model, the inclusive hadronic responses are given by the integration over the variables of the unobserved final nucleon and the summation over all initial nucleons:
\be
R_{L,T}= \int_0^{2\pi}{d\phi_N}\int_{-1}^1{d\cos{\theta_N}} \frac{K}{(2\pi)^3} \sum_\kappa {\cal R}^\kappa_{L,T}.
\ee
$\kappa$ represents the occupied nuclear shells (for neutrons and protons), $\theta_N$ and $\phi_N$ are the angles of the final nucleon, and $K$ is a function containing kinematical factors
\ba
K = \frac{M_B p_N M_N}{M_A f_{rec}} \hspace{1mm}, \hspace{1mm} f_{rec}=1+\frac{\omega p_N - q E_N \cos{\theta_N}}{M_A p_N} \,.\non\\
\ea
$M_A$ and $M_B$ are the masses of the target and residual system. The functions ${\cal R}^\kappa_{L,T}$ are the exclusive hadronic responses for each particular shell. They are linear combinations of different components of the hadronic tensor $H^{\mu\nu}_\kappa$:
\begin{align}
{\cal R}^{\kappa}_L &=\left(\frac{q^2}{Q^2}\right)^2 \left( H_{\kappa}^{00} - \frac{\omega}{q} (H_{\kappa}^{03}+H_{\kappa}^{30}) + \frac{\omega^2}{q^2} H_{\kappa}^{33} \right), \nonumber \\
{\cal R}^{\kappa}_T &= H_{\kappa}^{11} + H_{\kappa}^{22},
\end{align}
defined in a coordinate system with the $z$-axis in the direction of the transferred momentum ${\bf q}=(0,0,q)$. The hadronic tensor is given by
\be
    H_{\kappa}^{\mu\nu}= \sum_{m_j, s} [J_{\kappa,m_j,s}^\mu]^* J_{\kappa,m_j,s}^\nu.
\ee

The hadronic current, $J_{\kappa,m_j,s}^\mu$, includes all the processes that lead to a final 1p-1h state. It is computed between an initial $A$-body nuclear ground state and a final scattering state with a scattered nucleon and an $A-1$-body residual nucleus.  
We simplify the nuclear part by considering a shell model approach, introducing an independent particle description of the system in which each nucleon is subjected to a central potential created by the others. 
Then, the initial and final nuclear states can be expressed in terms of single-particle wave functions orthogonal to each other.
The final state is given by the product of a wave function for the $A-1$ residual nucleus and a distorted wave describing the ejected nucleon. In addition, factorizing the center of mass, the initial nuclear state is given by the product of an independent particle wave function coupled to the rest of the initial nucleus and the wave function for these remaining $A-1$ nucleons. Taking into account all these considerations, the hadronic current can be expressed in terms of the initial and final single-particle states, we finally get
\be
    J_{\kappa,m_j,s}^\mu =  \int{d{\bf p}} \Psib^{s}(\np_N', \np_N) \Gamma^\mu \Psi_\kappa^{m_j}({\bf p}),
    \label{eq:hadroniccurrent}
\ee
where $\np$ is the momentum of the bound nucleon and $m_j$ the third-component of its total angular momentum $j$. $\np_N$ is the asymptotic momentum of the final nucleon, $\np'_N$ its momentum inside the nucleus and $s$ its spin. The hadronic current operator $\Gamma^\mu$ includes all the processes that lead to a final 1p-1h state. 

The bound wave function $\Psi_\kappa^{m_j}$ is obtained using a RMF model which is an extension of the original $\sigma-\omega$ model including non-linear couplings to the $\sigma$ meson~\cite{Sharma93}. The starting point of the model is the construction of a phenomenological Lorentz covariant Lagrangian density that includes the nucleon-nucleon interaction through meson exchange. The parameters (couplings and mass of the $\sigma$ meson) are adjusted by fitting general properties of some finite nuclei.  
The bound wave function is obtained in the Hartree approach as a solution of the Dirac equation with spherically symmetric scalar and vector potentials, giving a 4-component spinor with good total angular momentum quantum numbers $\kappa$ and $m_j$,
\be
\Psi_\kappa^{m_j}({\bf p})=(-i)^l 
\begin{pmatrix}
g_\kappa(p) \\
S_\kappa f_\kappa(p) \frac{{\bs \sigma} \cdot \np}{p}
\end{pmatrix}
\varphi_\kappa^{m_j}(\Omega_p).
\label{eq:fdo-bound}
\ee

The distorted wave function of the knocked out nucleon $\Psi^s$ is obtained as a solution of the Dirac equation in the continuous with scalar and vector potentials. 
It is expressed as a partial wave expansion
\begin{multline}
\Psib^{s}({\bf p_N'}, \np_N)=4\pi \sqrt{\frac{E_N + M_N}{2M_N}} \sum_{\kappa,m_j,m} i^\ell e^{-i\delta_\kappa^*} \times \\ \times \langle \ell m \frac{1}{2} s_N |j m_j \rangle Y_\ell^{m*}(\hat{{\bf p}}_N) \phi_\kappa^{m_j}({\bf p_N'}),
\end{multline}
where $\phi_\kappa^{m_j}({\bf p_N'})$ are spinors as in eq.~\eqref{eq:fdo-bound}~\cite{Udias93,Martinez06,NikolakopoulosPhD}.
In this work, the final nucleon wave function  is described using the ED-RMF potential. 
This is a real potential that, by construction, is identical to the potential felt by the bound nucleons when there is no overlap between final and initial state, so that orthogonality is preserved. For increasing final nucleon momentum, when there is no overlap, the potential weakens, similarly as the usual phenomenological energy-dependent optical potentials (see details in \cite{Gonzalez-Jimenez19,Nikolakopoulos19, Gonzalez-Jimenez20}).

In the independent-particle shell model (IPSM), the nucleus $^{12}$C consists of 2 nucleons in the $1s_{1/2}$ state and 4 nucleons in the $1p_{3/2}$ state, and the missing energy distribution (referring to the portion of transferred energy that converts into internal energy of the residual nucleus) is the sum of two Dirac deltas. However, this simplistic approximation does not fully capture the actual missing energy distribution, which has been experimentally measured in $(e,e'p)$ experiments for carbon and other nuclei \cite{Dutta03,Fissum04}.
The experimental data reveals that the energy response of each shell exhibits a finite width. Additionally, it is observed that the experimental occupancy of the shells is reduced compared to the predictions of the IPSM. Correlations beyond the IPSM moves the nucleons from the independent-particle levels to deeper missing energy ($E_m$) and missing momentum ($p_m$) regions. 
To account for these effects, we incorporate into our formalism a more realistic missing energy profile inspired by the Rome spectral function~\cite{Benhar94,Benhar05}.
In particular, we decrease the occupation of the $1s_{1/2}$ and $1p_{3/2}$ shells to 3.3 and 1.8, respectively. Additionally, we consider the high missing energy and momentum region of the spectral function arising from short-range correlations, modeling it with an $s$-wave that is fitted to reproduce the momentum distribution of the Rome spectral function (for more details see \cite{Gonzalez-Jimenez22,Franco-Patino22}).

\section{Two-body contributions}\label{sec:two-body}

Now, we extend the usual treatment of QE scattering, based on a one-body current operator, and include one-pion exchange effects by incorporating two-body meson-exchange currents with a final paticle-hole state. Then, the hadronic current reads
\be
J^\mu_{\kappa,m_j,s}=J^\mu_{1b}+J^\mu_{2b}.
\label{eq:J}
\ee

The one-body part is given by the well-known expression
\be
J^\mu_{1b}=\int{d{\bf p}} \Psib^{s}(\np + {\bf q}, \np_N) \Gamma^\mu_{1b} \Psi_\kappa^{m_j}({\bf p}),
\ee
where the one-body current operator $\Gamma^\mu_{1b}$ is computed using the CC2 prescription \cite{Udias93,Udias95,Martinez06}. 

\subsection{The general case: intermediate RMF-nucleon approach}

\begin{figure}[htbp]
\centering  
\includegraphics[width=0.5\textwidth,angle=0]{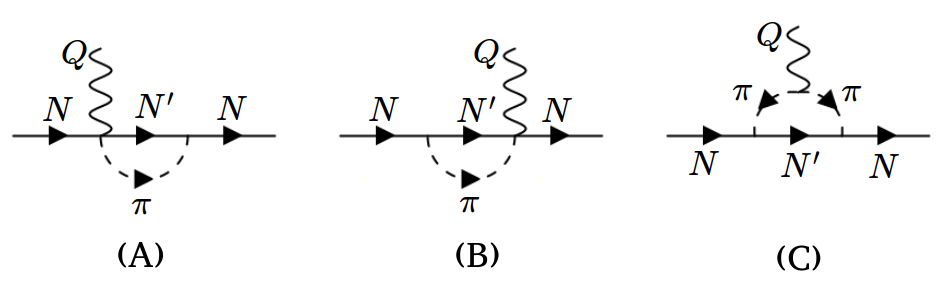}
\caption{Background contributions: seagull or contact [CT, (A) and (B)] and pion-in-flight [PF, (C)]. $N'$ denotes the intermediate bound-nucleon state.}
\label{fig:background}
\end{figure}

\begin{figure*}[htbp]
\centering  
\includegraphics[width=\textwidth,angle=0]{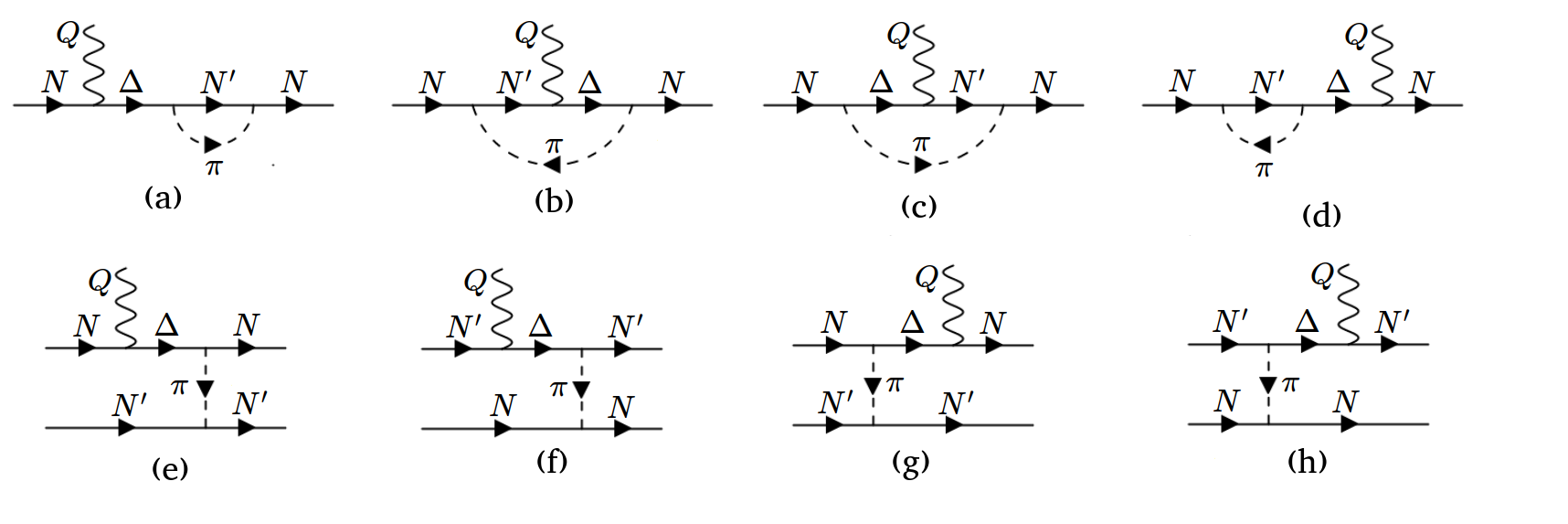}
\caption{Delta exchange [(a),(b),(c),(d)] and direct [(e),(f),(g),(h)] contributions. $N'$ denotes the intermediate bound-nucleon state. }
\label{fig:delta}
\end{figure*}

The two-body current is the sum of the contributions from the dominant Delta-resonance mechanism (diagrams in fig.~\ref{fig:delta}) and the background from the ChPT $\pi N$-Lagrangian~\footnote{ The expressions of the vertices used in this work can be found at Appendix A in \cite{Gonzalez-Jimenez17}.} (diagrams in fig. \ref{fig:background}). The particle-hole excitation occurs through a two-body current when, in the two-particle two-hole (2p-2h) interaction , one of the outgoing nucleons remains bound to the nucleus. In this way, the hadronic final state consists in just one scattered nucleon and there appears an intermediate bound-nucleon state, denoted as $N'$ in the diagrams.

Within the second quantization formalism, the general expression for any two-body operator is
\be
\hat{J}= \frac{1}{2} \sum_{\alpha_1, \alpha_{1'}, \alpha_2, \alpha_{2'}} c_{\alpha_{1'}}^\dagger c_{\alpha_{2'}}^\dagger c_{\alpha_{2}} c_{\alpha_{1}} J(\alpha_1, \alpha_{1'}, \alpha_2, \alpha_{2'})
\ee
where, denoting $F$ as the ground state of the target nucleus, 
\begin{align}
c_\alpha &=a_\alpha \hspace{2mm} \textnormal{ if }  \hspace{2mm}  \alpha>F \\ 
c_\alpha &=b^\dagger_\alpha  \hspace{2mm} \textnormal{ if }  \hspace{2mm}  \alpha<F
\end{align} 
with $a^\dagger_\alpha$ ($a_\alpha$) and $b^\dagger_\alpha$ ($b_\alpha$) the particle and hole creation (annihilation) operators, respectively. Holes are described by bound wave functions and particles by distorted wave functions. The subindex $\alpha$ represents the quantum numbers that label the single-particle states of the system, thus, they are different in each case.  For the holes, the quantum numbers are $\kappa$, $m_j$ and the isospin and, for the particles, the momenta, the spin and the isospin. For clearness, we omit the isospin subscripts.

We are interested in a particle-hole final state, in which a nucleon of the target nucleus in state $|\alpha \equiv \kappa, m_j \rangle$ is knocked out and detected in state $|\alpha_N \equiv P_N, s_N\rangle$,
\be
|\alpha_N; \alpha \rangle = a^\dagger_{\alpha_N} b^\dagger_\alpha |F \rangle. 
\ee
Then, the two-body hadronic current between the ground state and a particle-hole excited state reads
\begin{multline}
\langle \alpha_N; \alpha| \hat{J} | F \rangle =\\= \sum_{\alpha'<F} [J(\alpha, \alpha_N, \alpha', \alpha') - J(\alpha', \alpha_N,\alpha,\alpha')]
\label{eq:J2b-alpha}
\end{multline}
where antisymmetrization is implicitly included and the minus sign for fermionic loops is recovered in the resulting expression. $\alpha'$ denotes the quantum numbers of the intermediate bound-nucleon state and its summation runs over all occupied levels in the ground state. In the general case, the intermediate bound particles are described using the RMF model, so $\alpha'\equiv \kappa',m_j'$, and we refer to it as the intermediate RMF-nucleon approach.
An important difference with respect to the relativistic Fermi gas case used in other works \cite{Amaro02} is that the momentum is not a quantum number of RMF-bound states. Then, there is no restriction to the momentum of the intermediate nucleons to be the same. 
The terms $J(\alpha, \alpha_N, \alpha', \alpha')$ and  $J(\alpha', \alpha_N,\alpha,\alpha')$ represent the direct and exchange contributions, respectively. In the exchange terms, the 1p-1h excitation results when one of the outgoing nucleons of the 2p-2h excitation fills in the hole left by the other, so that it remains bound. In the direct terms, one of the final nucleons remains in its initial bound state after the interaction.  For the background diagrams, the direct terms vanishes due to the isospin dependence of the ChPT Lagrangian. Hence, only the exchange terms contribute (fig. \ref{fig:background}). On the other hand, the Delta-resonance part has contributions from both exchange and direct terms [respectively, diagrams (a-d) and (e-h) in fig.~\ref{fig:delta}].  For the delta case, every process can occur through intermediate proton or neutron, so both contributions have to be added.   

Finally, the two-body current reads,
\begin{multline}
  J^\mu_{2b} = \int{d{\bf p}} \int{\frac{d\np_p}{(2\pi)^{3/2}}} \int{\frac{d \np_h}{(2\pi)^{3/2}}} \\ \times \Psib^{s}(\np+\np_h+{\bf q}-\np_p, \np_N) \Gamma^\mu_{2b} \Psi_\kappa^{m_j}({\bf p}),  
  \label{eq:J2b}
\end{multline}
with 
$\np_p$ and $\np_h$ the momenta of the intermediate nucleons. The two-body operator is the sum of the Delta-resonance and background diagrams: $\Gamma^\mu_{2b}=\Gamma_{ChPT}^\mu + \Gamma_{\Delta}^\mu$.

The hadronic current operators for the background terms read 
\onecolumngrid
\begin{align}
\Gamma_{ChPT,(A)}^\mu &= I F_{CT} F_{\pi NN}^2   \frac{2f^2}{m_\pi^2} \slashed{K}_\pi \gamma^5 \frac{\Lambda_{N'}(\np_h,\np_p)}{K_\pi^2-m_\pi^2} \gamma^\mu \gamma^5, \\
\Gamma_{ChPT,(B)}^\mu &=  -I F_{CT} F_{\pi NN}^2  \frac{2f^2}{m_\pi^2} \gamma^\mu \gamma^5 \frac{\Lambda_{N'}(\np_h,\np_p)}{K_\pi^2-m_\pi^2} \slashed{K}_\pi \gamma^5, \\
\Gamma_{ChPT,(C)}^\mu &= I F_{PF} F_{\pi NN}(K^2_{\pi,1})  F_{\pi NN}(K^2_{\pi,2}) \frac{2f^2}{m_\pi^2} \frac{(Q+2P-2P_p)^\mu}{(K_{\pi,1}^2 - m_\pi^2) (K_{\pi,2}^2 - m_\pi^2)}  \slashed{K}_{\pi,1} \gamma^5  \Lambda_{N'}(\np_h,\np_p) \slashed{K}_{\pi,2} \gamma^5,
\end{align}
where $K_\pi^{(A)}=K_{\pi,1}^{(C)}=P+Q-P_p$ and $K_\pi^{(B)}=K_{\pi,2}^{(C)}=P_p-P$. For the Delta contribution, the current operators are given by
\begin{align}
\Gamma_{\Delta,(a)}^\mu &= - I F_{\pi NN} F_{\pi\Delta N } \frac{f}{m_\pi} 
 \slashed{K}_\pi \gamma^5 \frac{\Lambda_{N'}(\np_h,\np_p)}{K_\pi^2-m_\pi^2} \Gamma^\alpha_{\Delta \pi N'} S_{\Delta,\alpha \beta} \Gamma_{\gamma\Delta N}^{\beta \mu}, &&P_{\Delta,a}= P+Q \\
\Gamma_{\Delta,(b)}^\mu&= - I  F_{\pi NN} F_{\pi\Delta N }  \frac{f}{m_\pi}  \Gamma^\alpha_{\Delta \pi N} S_{\Delta,\alpha \beta} \Gamma_{\gamma\Delta N'}^{\beta \mu}  \frac{\Lambda_{N'}(\np_h,\np_p)}{K_\pi^2-m_\pi^2} \slashed{K}_\pi \gamma^5 ,   &&P_{\Delta,b} = Q+P_h \\
\Gamma_{\Delta,(c)}^\mu &= - I F_{\pi NN} F_{\pi\Delta N } \frac{f}{m_\pi}   \slashed{K}_\pi \gamma^5 \frac{\Lambda_{N'}(\np_h,\np_p)}{K_\pi^2-m_\pi^2}   \bar{\Gamma}^{\alpha\mu}_{\gamma \Delta N'} S_{\Delta,\alpha \beta} \Gamma_{\Delta \pi N}^{\beta}, &&P_{\Delta,c}=P_p-Q \\
\Gamma_{\Delta,(d)}^\mu &= - I F_{\pi NN} F_{\pi\Delta N } \frac{f}{m_\pi}  \bar{\Gamma}^{\alpha\mu}_{\gamma \Delta N} S_{\Delta,\alpha \beta} \Gamma_{\Delta \pi N'}^{\beta} \frac{\Lambda_{N'}(\np_h,\np_p)}{K_\pi^2-m_\pi^2} \slashed{K}_\pi \gamma^5 , &&P_{\Delta,d}=P_N'-Q \\
\Gamma_{\Delta,(e)}^\mu &= I F_{\pi NN} F_{\pi\Delta N } \frac{f}{m_\pi} 
 \frac{\Lambda_{\pi N'}(\np_h,\np_p,K_\pi)}{K_\pi^2-m_\pi^2}  \Gamma^\alpha_{\Delta \pi N} S_{\Delta,\alpha \beta} \Gamma_{\gamma\Delta N}^{\beta \mu} ,   &&P_{\Delta,e}= P+Q  \\
\Gamma_{\Delta,(f)}^\mu &= I  F_{\pi NN} F_{\pi\Delta N } \frac{f}{m_\pi} \frac{\Lambda_{\Delta N'}(\np_h,\np_p,P_{\Delta,f})}{K_\pi^2-m_\pi^2}  \slashed{K}_\pi \gamma^5, &&P_{\Delta,f}= Q + P_h \\
\Gamma_{\Delta,(g)}^\mu &= I F_{\pi NN} F_{\pi\Delta N } \frac{f}{m_\pi} \frac{\Lambda_{\pi N'}(\np_h,\np_p,K_\pi)}{K_\pi^2-m_\pi^2}  
\bar{\Gamma}^{\alpha\mu}_{\gamma \Delta N} S_{\Delta,\alpha \beta} \Gamma_{\Delta \pi N}^{\beta}, &&P_{\Delta,g}= P_N'-Q \\
\Gamma_{\Delta,(h)}^\mu &= I  F_{\pi NN} F_{\pi\Delta N } \frac{f}{m_\pi} \frac{\bar{\Lambda}_{\Delta N'}(\np_h,\np_p,P_{\Delta,h})}{K_\pi^2-m_\pi^2}  \slashed{K}_\pi \gamma^5 , &&P_{\Delta,h}=P_p-Q
\end{align}
where $K_\pi^{(a)}=K_\pi^{(c)}=P+Q-P_p$, $K_\pi^{(b)}=K_\pi^{(d)}=P_p-P$, $K_\pi^{(e)}=K_\pi^{(g)}=P+Q-P_N'$, $K_\pi^{(f)}=K_\pi^{(h)}=Q+P_h-P_p$ and $\bar{\Gamma}^{\alpha \mu}_{\gamma \Delta N}(P^\mu,Q^\mu)=\gamma^0 \left[\Gamma^{\alpha \mu}_{\gamma \Delta N}(P^\mu,-Q^\mu)\right]^\dagger \gamma^0$. $I$ is the isospin coefficient of each diagram, given in Table \ref{table:isospin}, and to shorten the expressions we have introduced the intermediate RMF projectors,
\begin{align}
\Lambda_{N'}(\np,\np') &= \sum_{\kappa',m_j'} \Psi_{\kappa'}^{m_j'}(\np) \Psib_{\kappa'}^{m_j'}(\np'), \\
\Lambda_{\pi N'}(\np, \np',K_\pi) &=\sum_{\kappa',m_j'}  \Psib_{\kappa'}^{m_j'}(\np') \slashed{K}_\pi \gamma^5 \Psi_{\kappa'}^{m_j'}(\np), \\
\Lambda_{\Delta N'}(\np, \np',P_\Delta) &=\sum_{\kappa',m_j'} \Psib_{\kappa'}^{m_j'}(\np')
 \Gamma^\alpha_{\Delta \pi N'} S_{\Delta,\alpha \beta} \Gamma_{\gamma\Delta N'}^{\beta \mu} \Psi_{\kappa'}^{m_j'}(\np), \\
\bar{\Lambda}_{\Delta N}(\np, \np',P_\Delta) &=\sum_{\kappa',m_j'} \Psib_{\kappa'}^{m_j'}(\np')
 \bar{\Gamma}^{\alpha\mu}_{\gamma \Delta N'} S_{\Delta,\alpha \beta} \Gamma_{\Delta \pi N'}^{\beta} \Psi_{\kappa'}^{m_j'}(\np).
 \end{align}
 \twocolumngrid
Finally, to account for the nucleon structure we introduce form factors in the background operators,
\be
F_{CT}(Q^2)=F_{PF}(Q^2)=F_1^V(Q^2),
\ee
where $F_1^V$ is the isovector nucleon form factor. Also, we add a strong form factor in the $\gamma \pi NN$ and $\pi NN$ vertices, $F_{\pi NN}$, and in the $\pi \Delta N$ vertex, $F_{\pi \Delta N}$,  which accounts for the off-shell nature of the pion: 
\be
F_{\pi NN}(K^2_\pi)=\frac{\Lambda^2-m_\pi^2}{\Lambda^2-K_\pi^2}, \hspace{2mm} F_{\pi \Delta N}(K^2_\pi)=\frac{\Lambda^2_{\pi\Delta N}}{\Lambda^2_{\pi\Delta N}-K_\pi^2},
\ee
with $\Lambda=1.3$ GeV \cite{Amaro02,Dekker94} and $\Lambda^2_{\pi \Delta N}=1.5 M_N^2$ \cite{DePace03,Dekker94}.

\begin{table}[h]
\begin{center}
\begin{tabular}{ c | c c c c}
Channel & CT & PF & $\Delta$ (a, d, e, g) & $\Delta$ (b, c, f, h) \\ \hline
p $\rightarrow$ p (N'=p)& 0 & 0 & $1/\sqrt{3}$ & $1/\sqrt{3}$ \\
p $\rightarrow$ p (N'=n)& 1 & 1 & $-1/\sqrt{3}$ & $1/\sqrt{3}$ \\
n $\rightarrow$ n (N'=p) & -1 & -1 & $1/\sqrt{3}$ & $-1/\sqrt{3}$\\
n $\rightarrow$ n (N'=n) & 0 & 0 & $-1/\sqrt{3}$ & $-1/\sqrt{3}$
\end{tabular}
\caption{Isospin coefficients $I$ for the possible contributions to meson exchange currents. $N'$ denotes the intermediate bound-nucleon state. CT and PF refer to the background diagrams in figure \ref{fig:background}, while $\Delta$ (a-h) refers to the delta diagrams in figure \ref{fig:delta}.}
\label{table:isospin}
\end{center}
\end{table}

\subsection{The intermediate RFG-nucleon approximation}

The complexity of these expressions, in particular the appearance of a 9-dimensional integral in the two-body current \eqref{eq:J2b}, makes the computational time extremely high.  For this reason, it is useful to describe the intermediate bound-nucleon state as free Dirac spinors in an RFG, in the same way as done in infinite nuclear matter \cite{Amaro02}. Then, the summation over the occupied levels of the ground state in eq.~\eqref{eq:J2b-alpha} now implies a sum over the intermediate momentum $\np_{ph}$, spin and isospin. In contrast with the RMF-nucleon case discussed above, here one has the constraint that the momentum of the intermediate nucleons must be the same and the hadronic current is reduced to a 6-dimensional integral. Additionally, in an isospin symmetric nucleus, 
the Delta-resonance direct terms and the exchange diagrams (a) and (d) vanish due to the sum over spin and isospin. 

Under this approximation, the two-body current can be written as
\begin{multline}
    J^\mu_{2b,free}= \int{d\np} \int{\frac{d\np_{ph}}{(2\pi)^{3}}} \Theta(p_F - p_{ph})  \\ \times \Psib^{s}(\np + {\bf q}, \np_N) \Gamma^\mu_{2b,free} \Psi_\kappa^{m_j}({\bf p}) 
\end{multline}
with
\begin{align}
  \Gamma^\mu_{2b,free} &= \Gamma_{ChPT,(A)}^\mu + \Gamma_{ChPT,(B)}^\mu + \Gamma_{ChPT,(C)}^\mu \nonumber \\ &+ \Gamma_{\Delta,(b)}^\mu + \Gamma_{\Delta,(c)}^\mu.  
\end{align}
Note that, now, we can reorganize the expression of the two-body current and write the hadronic current including one- and two-body contributions [eq. \eqref{eq:J}] as
\be
J^\mu_{\kappa,m_j,s}=\int{d\np} \Psib^{s}(\np + {\bf q}, \np_N) \Gamma^\mu \Psi_\kappa^{m_j}({\bf p}),
\ee
with
\be
\Gamma^\mu = \Gamma_{1b}^\mu + \int{\frac{d\np_{ph}}{(2\pi)^{3}}} \Theta(p_F - p_{ph}) \Gamma^\mu_{2b,free}.\label{hc-1b+2b}
\ee
In \eqref{hc-1b+2b} $p_F$ is the Fermi momentum of the nucleus, for carbon 12 we use the usual value 228 MeV, both for protons and neutrons.

Within this approximation, the hadronic current operators for the background terms are given by
\onecolumngrid
\begin{align}
    \Gamma_{ChPT,(A)}^\mu &= I F_{CT} F_{\pi NN}^2 \frac{2f^2}{m_\pi^2}  \frac{M}{E_{ph}} \slashed{K}_\pi \gamma^5 \frac{\Lambda(P_{ph})}{K_\pi^2-m_\pi^2} \gamma^\mu \gamma^5,
    \label{eq:MECa}
\\
    \Gamma_{ChPT,(B)}^\mu &= -I F_{CT} F_{\pi NN}^2  \frac{2f^2}{m_\pi^2}  \frac{M}{E_{ph}} \gamma^\mu \gamma^5 \frac{\Lambda(P_{ph})}{K_\pi^2-m_\pi^2} \slashed{K}_\pi \gamma^5 ,
    \label{eq:MECb}
\\
    \Gamma_{ChPT,(C)}^\mu &= I F_{PF} F_{\pi NN}(K^2_{\pi,1}) F_{\pi NN}(K^2_{\pi,2}) \frac{2f^2}{m_\pi^2}   \frac{M}{E_{ph}}  \frac{(Q+2P-2P_{ph})^\mu}{(K_{\pi,1}^2-m_\pi^2)(K_{\pi,2}^2-m_\pi^2)}  \slashed{K}_{\pi,1} \gamma^5 \Lambda(P_{ph}) \slashed{K}_{\pi,2} \gamma^5,
    \label{eq:MECc}
\end{align}
with $K_\pi^{(A)}=K_{\pi,1}^{(C)}=P+Q-P_{ph}$ and $K_\pi^{(B)}=K_{\pi,2}^{(C)}=P_{ph}-P$. For the contributing Delta-resonance terms, the hadronic current operators reads
\begin{align}
    \Gamma_{\Delta,(b)}^\mu &= - I F_{\pi N N} F_{\pi \Delta N} \frac{f}{m_\pi}  \frac{M}{E_{ph}} \Gamma_{\Delta \pi N}^\alpha  S_{\Delta,\alpha \beta} \Gamma^{\beta \mu}_{\gamma \Delta N}  \frac{\Lambda(P_{ph})}{K_\pi^2-m_\pi^2}\slashed{K}_\pi \gamma^5, &&P_\Delta^{(b)}=P_{ph}+Q
\label{eq:Deltab}
\\
    \Gamma_{\Delta,(c)}^\mu &= - I F_{\pi N N} F_{\pi \Delta N}   \frac{f}{m_\pi}  \frac{M}{E_{ph}} \slashed{K}_\pi \gamma^5  \frac{\Lambda(P_{ph})}{K_\pi^2-m_\pi^2} \bar{\Gamma}^{\alpha \mu}_{\gamma \Delta N}  S_{\Delta,\alpha \beta} \Gamma_{\Delta \pi N}^\beta, && P_\Delta^{(c)}=P_{ph}-Q
\label{eq:Deltac}
\end{align}
\twocolumngrid
with $K_\pi^{(b)}=P_{ph}-P$ and $K_\pi^{(c)}=P+Q-P_{ph}$.  As before, $I$ is the isospin coefficient, given in Table \ref{table:isospin}, and to shorten the expressions we have introduced the intermediate RFG projector
\ba
 \Lambda(P_{ph}) = \frac{\slashed{P}_{ph}+M}{2M}.
\ea
In this case, the projector is the same for protons and neutrons. 

\subsection{The modified intermediate RFG-nucleon approximation} 

In this approach we extend the RFG-nucleon approximation by including the relativistic interaction of nucleons through a modified energy and mass due to the scalar and vector potentials. The attractive scalar potential is accounted for in the relativistic effective mass
\be
M^*=m^* M < M.
\ee
Meanwhile, the vector potential produces a repulsive energy, which is added to the on-shell energy to obtain the modified nucleon energy
\be
E^*=E + E_v,
\ee
where $E=\sqrt{p^2 + (M^*)^2}$ is the on-shell energy with effective mass $M^*$. Then, the responses are computed as in the intermediate RFG-nucleon case with the change $M \rightarrow M^*$ and $E_{ph} \rightarrow E_{ph}^*$ in the intermediate nucleon variables. 
The $\Delta \gamma N$ vertex remains with the unmodified mass \cite{Amaro21}.

\begin{figure}[htbp]
\centering  
\includegraphics[width=0.48\textwidth,angle=0]{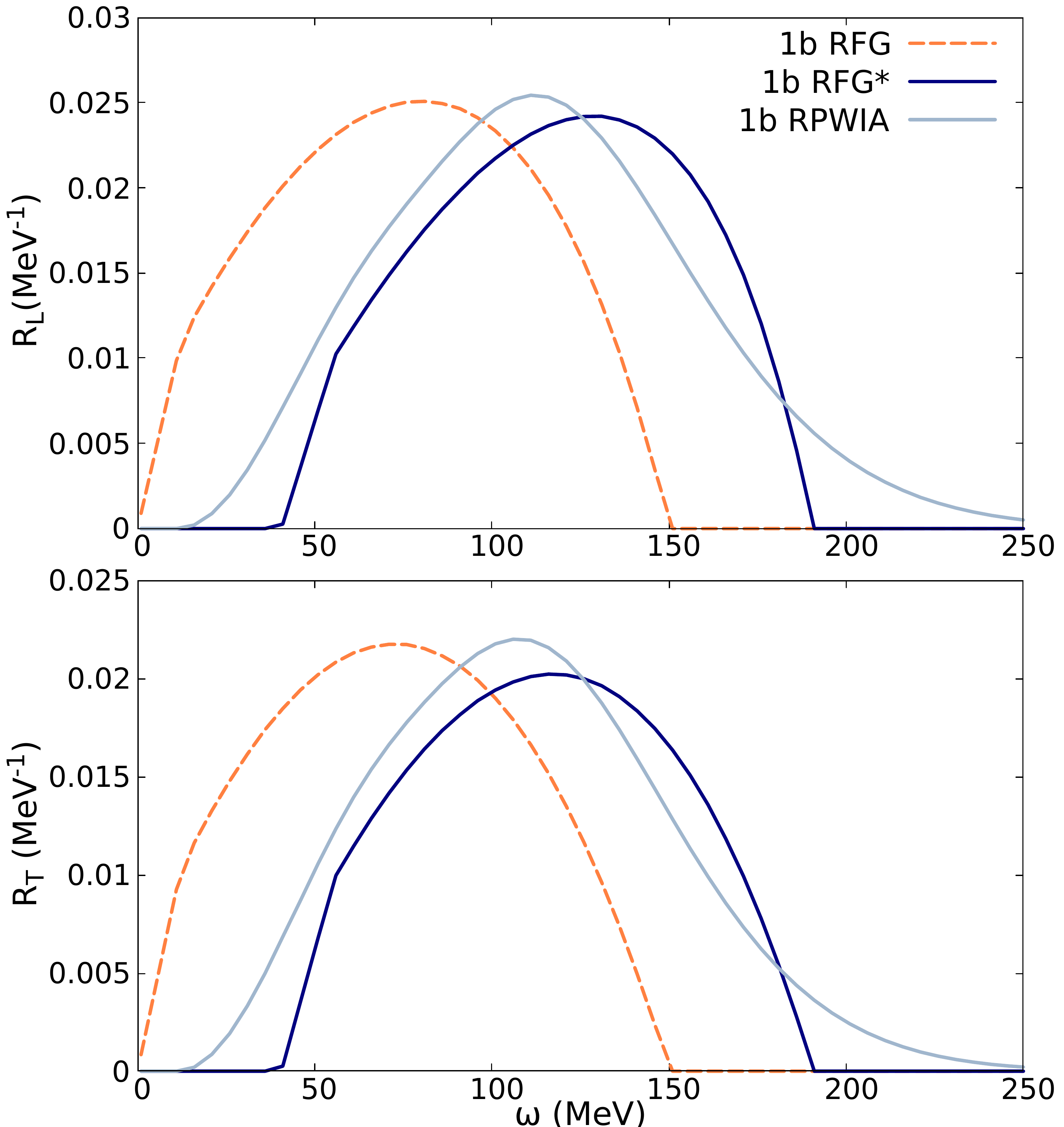}
\caption{$^{12}$C longitudinal (up) and transverse (bottom) electromagnetic inclusive response functions considering only one-body currents. The transferred momentum $q$ is 380 MeV/$c$. We show results for the relativistic Fermi gas (RFG), the relativistic Fermi gas with a modified initial nucleon (RFG$^*$) and  the relativistic plane wave impulse approximation (RPWIA).}
\label{fig:1b-RFG*}
\end{figure}

Following reference \cite{Amaro21}, we use an effective mass with $m^*=0.8$  and a vector energy of $E_v=141$ MeV for $^{12}$C. 
To validate this approach, in figure \ref{fig:1b-RFG*}, we compare the inclusive responses from the relativistic plane wave impulse approximation (RPWIA) and from a relativistic Fermi gas with a modified initial nucleon (RFG$^*$), both computed with only one-body currents. In the RPWIA model the initial nucleon is described by a bound wave function, so one would expect the results of RFG$^*$ and RPWIA to be similar. Indeed, one observes that modifying the mass and energy of the initial nucleon in the RFG produces a shift of the response to higher transferred energy and a slight decrease of the strength, looking more like the RPWIA case. Motivated by these results~\footnote{Though not shown here, we have made this study for several $q$, obtaining similar outcomes.}, we have applied this approach to our study of the two-body contributions.  

In what follows, we will refer to this approach as intermediate RFG*-nucleon approximation. It has been employed in the calculations of reference~\cite{Franco-Munoz22}.

%% file: 3_results.tex
\section{Results}\label{sec:results}

\begin{figure}[htbp]
\centering  
\includegraphics[width=0.48\textwidth,angle=0]{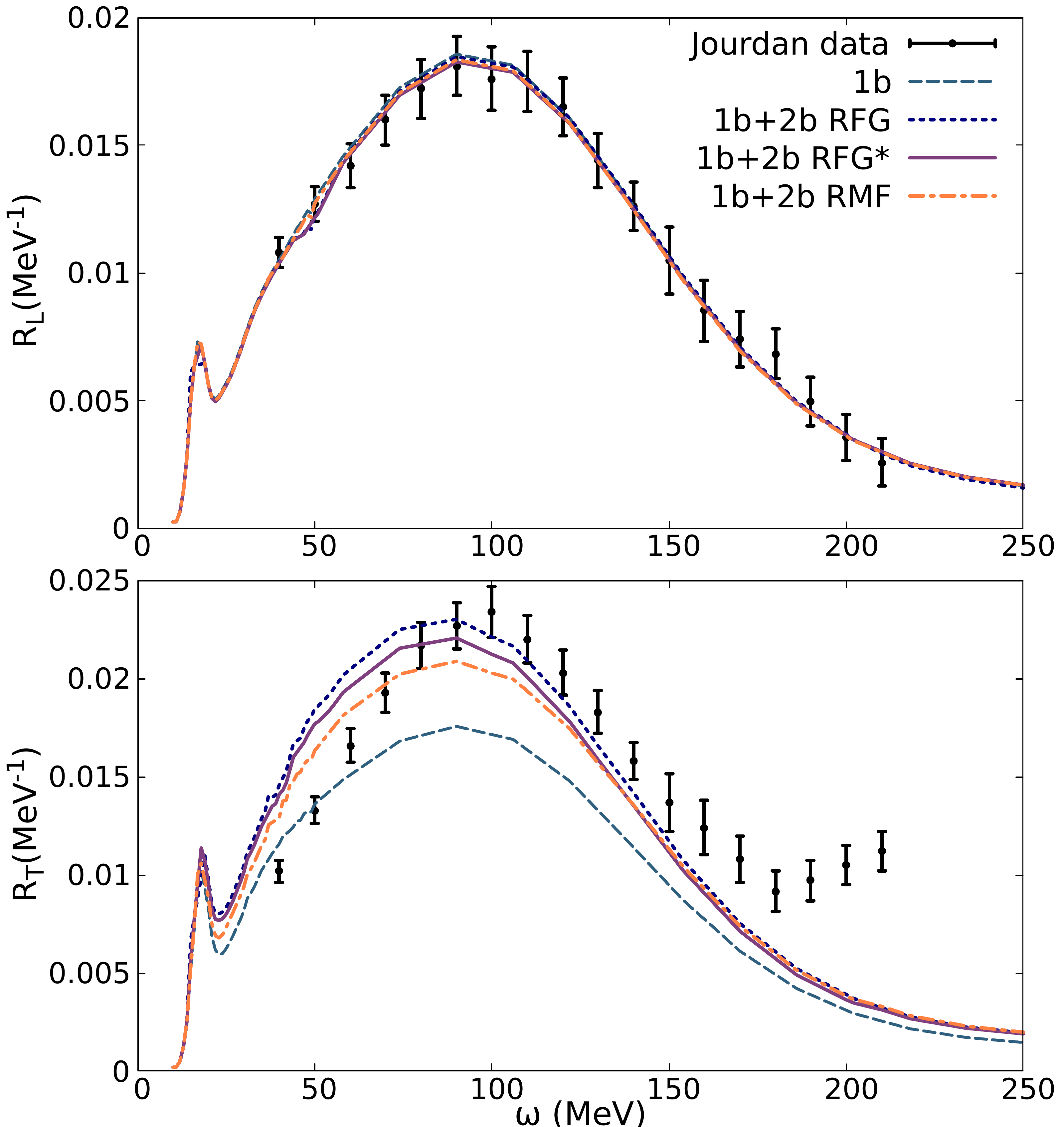}
\caption{$^{12}$C longitudinal (up) and transverse (bottom) electromagnetic inclusive response functions. The transferred momentum $q$ is 380 MeV/$c$. We show our results when the intermediate bound-nucleon state is described in terms of free particles in an RFG, including a modified mass and energy (RFG*), and RMF nucleons. Data are from Jourdan \cite{Jourdan96}.}
\label{fig:comparison}
\end{figure}

In figure~\ref{fig:comparison} we show the comparison of the different approaches that can be taken to describe the intermediate bound-nucleon state: the simplest case, RFG nucleons, its extension including scalar and vector potentials and, finally, the most complete approach with intermediate RMF nucleons. Our predictions  of the $^{12}$C  electromagnetic inclusive responses are compared to data extracted by Jourdan \cite{Jourdan96} by means of a Rosenbluth separation. 
In general terms, the two-body current operator results in an increase of the transverse response and a tiny effect on the longitudinal sector. 
The more realistic the treatment of the intermediate bound-nucleon state, the lower the increase. This gives rise to an increase of the transverse response up to 31\%, 25\% and 19\% for the intermediate RFG-nucleon case, its extension including scalar and vector potentials and the RMF one, respectively. The agreement of our results with data is outstanding for the longitudinal channel, and improved for the transverse one with the introduction of the two body currents. 

\begin{figure*}[ht!]
\centering  
\includegraphics[width=0.48\textwidth,angle=0]{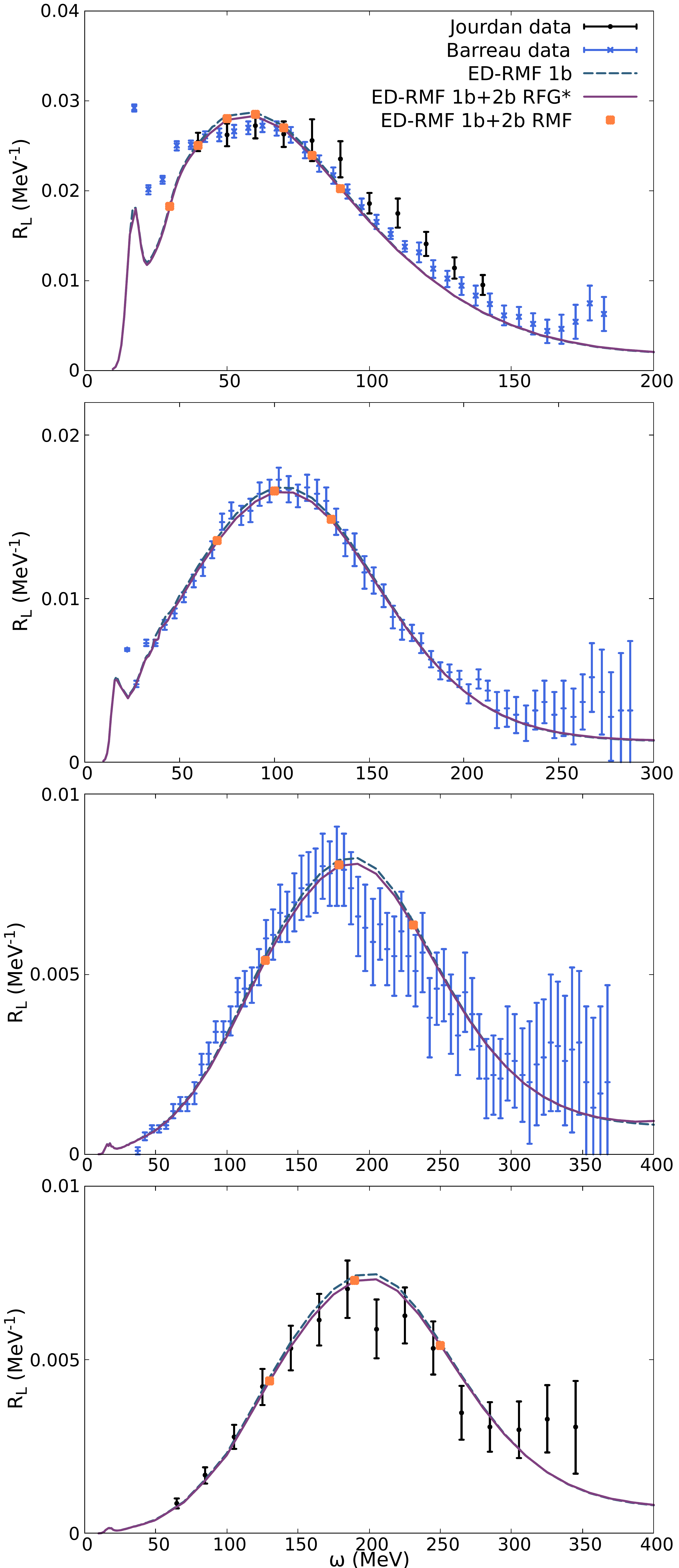}
\includegraphics[width=0.48\textwidth,angle=0]{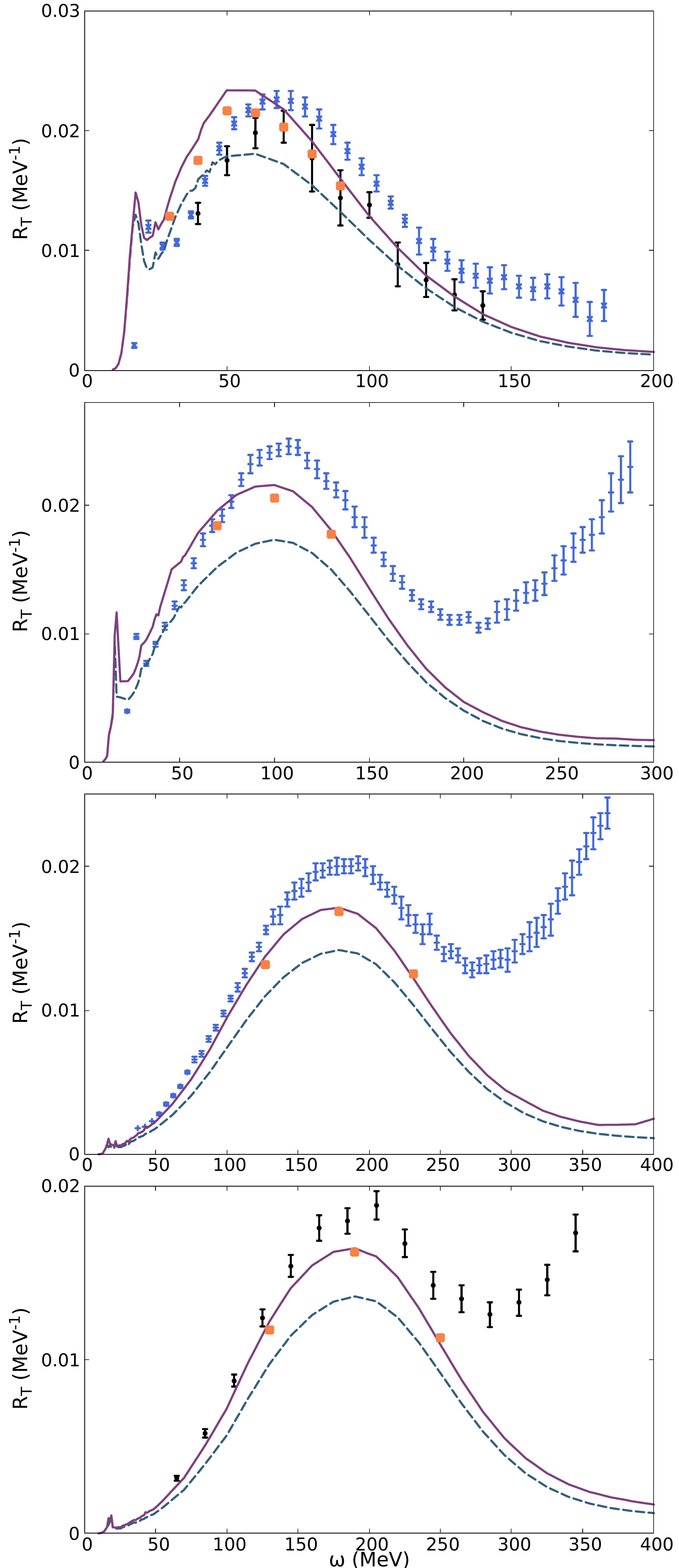}
\caption{$^{12}$C longitudinal (left) and transverse (right) responses. The transferred momentum $q$ is (from up to bottom) 300, 400, 550 and 570 MeV/$c$. We show our results when the intermediate bound-nucleon state is described in terms of free particles in an RFG, including a modified mass and energy (RFG*), and RMF nucleons. Data are from Jourdan \cite{Jourdan96} and Barreau et al. \cite{Barreau83}. }
\label{fig:RL-RT}
\end{figure*}

In figure \ref{fig:RL-RT} we show our results for the inclusive longitudinal and transverse responses for four different kinematics, computed using the one- and two-body operators and the ED-RMF potential to describe the final nucleon. We show the responses computed within the intermediate RFG*- and RMF-nucleon approaches, but given the extremely high computational cost only a few points are shown for the latter. The difference between the two approaches is smaller for larger values of $q$, obtaining essentially identical results for momentum transfer around and above $500$ MeV/$c$. 
This fact motivated the choice of the intermediate RFG*-nucleon approximation in the calculations of reference~\cite{Franco-Munoz22}. 
Meanwhile, at low $q$, the RMF description of the intermediate nucleons reduces the transverse increase, especially, at low energy transfer.

As mentioned, the intermediate RFG*-nucleon approximation has the advantage of reducing the two-body current from a 9- to a 6-dimensional integral as well as less contributing diagrams, in contrast with the RMF case. 
While it is possible to compute the $^{12}$C inclusive responses in a manageable amount of time for the RFG* approach, the computational effort required for the RMF one makes it impractical for its use in predictions of neutrino-nucleus cross sections, where one has to average over the neutrino flux. 
Luckily, in most of accelerator-based neutrino experiments, the neutrino energy centers at around 1 GeV or above, and for those energies most of the strength of the cross section comes from $q>500$ MeV/$c$~\cite{Megias14}. Therefore, the RFG* approach would be an excellent approximation to the complete model. 

Finally, we point out that it is expected to underestimate the inclusive data, especially in the high energy transfer region, because other processes, as 2p-2h and pion production, contribute. 

%% file: 4_conclusions.tex
\section{Conclusions}\label{sec:conclusions}

The main contribution of this work is the development of a model for the computation of the particle-hole electromagnetic responses of $^{12}$C that includes two-body meson exchange currents. It incorporates the contribution from the Delta-resonance mechanism and the background from a ChPT $\pi N$-Lagrangian. The introduction of two-body currents results in an increase of the transverse response, improving the agreement with the data, meanwhile the longitudinal part remains practically unchanged. 

We have presented different approaches to describe the intermediate bound-nucleon state which appears in the particle-hole excitation through two-body currents.  Our study considers a shell model description of the nuclear structure. Then, we start the computation describing the intermediate nucleons as bound wave functions using the same RMF potential as that of the initial and final nucleons. The key point of this approach is that orthogonality is preserved between all particle states. However, the hadronic current implies the computation of a 9-dimensional integral with the contribution of a high number of diagrams, requiring extremely high computational times. For this reason, two possible approximations that simplify the problem have been presented.
First, we approximated the intermediate bound-nucleon state by free Dirac spinors in an RFG, reducing significantly the computational effort as the hadronic current is reduced to a 6-dimensional integral with less diagrams contributing. However, this approach leads to the lost of consistency between the intermediate state and the initial and final ones, as well as the bound aspect of the nucleons.
To account for this bound condition, we introduce a modified energy and mass due to the mean-field scalar and vector potentials. 
The differences with respect to the complete RMF case are notably reduced. 
This allows us to estimate the inclusive responses of $^{12}$C in a feasible amount of time for a variety of kinematics. Finally, the transverse response show an increase with respect to the one-body current result of around 31\% and 21\% for the RFG* and RMF approaches, respectively.

The next step will be to apply the present model to heavier nuclei, in particular, we will focus on argon 40, which is of great interest for the neutrino oscillation experiments MicroBooNE and DUNE. On the other hand, after this first validation of our theoretical model with the electron scattering case, we will explore the neutrino-nucleus interaction.